\documentclass[numbook,epj]{svjour}

\usepackage{stmaryrd,color}
\usepackage{epsfig}
\usepackage{amsmath}
\usepackage{graphicx}
\usepackage{dcolumn}
\usepackage{bm}
\usepackage{amssymb}
\usepackage{amsmath}
\usepackage{epsf}
\usepackage{subfigure}
\usepackage{epstopdf}
\usepackage{wrapfig}
\usepackage{pifont}
\usepackage{color}
\usepackage{wasysym}
\usepackage{slashed}
\usepackage{hyperref}
\usepackage{appendix}
\usepackage{subfigure}
\usepackage{csquotes}
\everymath{\displaystyle}

\begin{document}
\title{Generalized Drude Scattering rate from the memory function formalism: an independent verification of the Sharapov-Carbotte result}

\author{Pankaj Bhalla\inst{1,2} \and Navinder Singh\inst{1}
}                     
%
\mail{pankaj@prl.res.in}        
\institute{Physical Research Laboratory, Navrangpura, Ahmedabad-380009 India \and Indian Institute of Technology Gandhinagar-382424, India}
\date{Received: date / Revised version: date}

\abstract{
An explicit perturbative computation of the Mori's memory function was performed by G\"otze and W\"olfle (GW) to calculate Generalized Drude scattering (GDS) rate for the case of electron-impurity and electron-phonon scattering in metals by assuming constant electronic density of states at the Fermi energy. In the present investigation, we go beyond this assumption and extend the GW formalism to the case in which there is a gap around the Fermi surface in electron density of states. The resulting GDS is compared with a recent one by Sharapov and Carbotte (SC) obtained through a different route. We find good agreement between the two at finite frequencies. However, we find discrepancies in the dc scattering rate. These are due to a crucial assumption made in SC namely $\omega >> \vert \Sigma(\epsilon+\omega) - \Sigma^{*}(\epsilon)\vert$. No such high frequency assumption is made in the memory function based technique. 
\PACS{
      {74.25.Gz}{Optical properties}   \and
      {72.10.-d}{Theory of electronic transport; scattering mechanisms}
     } 
} 
\maketitle
\section{Introduction}
\label{intro}
The study of transport properties like optical conductivity is very important to understand the electronic interactions in complex many body systems like cuprates\cite{basov_05,basov_11}. The electronic interactions comprises of electron-phonon, electron-boson (spin-fluctuations), electron-impurity, elec-tron-electron interactions. Experimentally, the signatures of these interactions can be grasped by using optical data ($\sigma(\omega,T)$)\cite{timusk_03,puchkov_96} which includes the deduction of the Generalized Drude scattering (GDS) rate and mass enhancement factor using the standard form \begin{equation}\sigma(\omega,T)=\frac{\omega_{p}^{2}}{4\pi}\frac{1}{1/\tau(\omega,T)+i\omega(1+\lambda(\omega,T))}.
\end{equation}
Here $1/\tau(\omega,T)$ is the frequency and temperature dependent scattering rate, $\lambda(\omega,T)$ is the frequency and temperature dependent mass enhancement factor and $\omega_{p}$ is the plasma frequency. On theory side, the derivation of the analytical formulae for these quantities (like $1/\tau(\omega,T)$ and $\lambda(\omega,T)$) is very complicated.

Generally, the optical conductivity is calculated using Boltzmann's equation by assuming relaxation time approximation\cite{ziman_60,pines_96,ashcroft_76}. In that picture, scattering rate is considered as constant (independent of frequency and temperature) and the resulting behaviour corresponds to the Drude behaviour. In systems such as cuprates where the electron-boson, electron-electron interactions are important, this approach is inadequate as experiments show that scattering rate {\it{does}} depend on frequency and temperature \cite{dai_12,lee_05}. The complete solution of the transport problem in cuprates and other strongly correlated materials is complicated as there are no controlled perturbation parameters. For example, in simple metals where, as first shown by Holstein\cite{holstein_64}, the perturbation parameter is $v_{s}/v_{F}$ (sound speed/Fermi velocity) which is a small parameter and perturbative calculations are justified. Building upon Holstein's work on metals\cite{holstein_64}, Allen\cite{allen_71} has derived a relation for frequency dependent scattering rate. This has been further generalized by Mitrovi\'c and Fiorucci\cite{mitrovic_85} by considering the effects of non-constant density of states. Further this has been extended recently for the finite temperature case by Sharapov and Carbotte\cite{sharapov_05}. All these approaches are based on some assumptions such as neglecting vertex corrections. To go beyond this assumption, we have developed a formula for frequency and temperature dependent scattering rate using memory function technique\cite{mori_65,gotze_71} which includes the effect of vertex corrections\cite{gotze_72} (where the current-current correlators are directly computed without writing them in terms of single-particle Green's function). This technique is a generalization of Zwanzig projection operator technique\cite{zwanzig_61a,zwanzig_61}.  Physically, this approach is very appealing, because the conductivity $\sigma(\omega,T)$ can be cast into the generalized Drude form with frequency and temperature dependent scattering rate. Recently, it has also been used by several authors to study the transport properties of different systems\cite{plakida_96,plakida_97,vladimirov_12,forster_95,fulde_12,lucas_15,subir_15,subir_14,nabyendu_15}.

The paper is organized as follows. In Sec.II, we have elaborated the G\"otze-W\"olfle memory function formalism \cite{gotze_72}. In Sec.III, we go beyond the constant electronic density of states assumption and introduce the gapped density of states and calculate the imaginary part of memory function. Here we also discuss the dc and ac imaginary part of memory function in different temperature regimes and in appropriate limits we reproduce GW results. In Sec.IV, we compare our findings with SC result\cite{sharapov_05} and finally we conclude with a brief discussion in Sec.V.
\section{G\"otze-W\"olfle Formalism for electron-phonon scattering}
\label{sec:1}
In this section, a short introduction to GW formalism is presented\cite{gotze_72}. The Hamiltonian used for electron-phonon interaction is given by
\begin{equation}
H=H_{0}+H_{\textnormal{ep}}+H_{\textnormal{p}},
\label{ham}
\end{equation}
where $H_{0}$ is the Hamiltonian for \textit{non-interacting electrons} and is represented as
\begin{equation}
H_{0}=\sum_{\textbf{k}, \sigma} \epsilon(\textbf{k})c_{\textbf{k},\sigma}^{\dagger}c_{\textbf{k},\sigma},
\end{equation}
where $\epsilon(\textbf{k})$ is the band dispersion and $c_{\textbf{k},\sigma}^{\dagger}$, $c_{\textbf{k},\sigma}$ are creation and annihilation operators with wave vector $\textbf{k}$ and spin $\sigma$. The Hamiltonian $H_{\textnormal{ep}}$ represents the electron - phonon interaction and is given by
\begin{equation}
H_{\textnormal{ep}}=\sum_{\textbf{k},\textbf{k}^{\prime},\sigma} \left[ D(\textbf{k}-\textbf{k}') c_{\textbf{k}^{\prime}, \sigma}^{\dagger} c_{\textbf{k}, \sigma} b_{\textbf{k}-\textbf{k}^{\prime}} + \textnormal{h.c.} \right].
\label{eb}
\end{equation}
Here $b_{\textbf{k}-\textbf{k}'}$, $b_{\textbf{k}-\textbf{k}'}^{\dagger}$ are the annihilation and creation operators for phonons and $D(\textbf{k}-\textbf{k}')$ is the electron-phonon matrix element. Here the symbol $h.c.$ corresponds to the Hermitian conjugate of first term. The third part of eqn.(\ref{ham}) represents the free phonon Hamiltonian
\begin{equation}
H_{\textnormal{p}}=\sum_{q} \omega_{q} \left( b_{q}^{\dagger} b_{q} +\frac{1}{2} \right),
\end{equation} 
where $\omega_{q}$ is the phonon frequency.

According to the linear response theory, the dynamical conductivity is defined as\cite{kadanoff_63,zubarev_60,mahan_90,arfi_92}
\begin{equation}
\sigma(z)=-i \frac{1}{z} \chi(z) + i \frac{\omega_{p}^{2}}{4\pi z}.
\label{cond}
\end{equation}
Here $\omega_{p}^{2}=4\pi N_{e} e^{2} /m$ is the square of plasma frequency where $e$ electronic charge, $m$ electron mass and $N_{e}$ is the electron density, $z$ is the complex frequency and $\chi(z)$ is the current-current correlation function defined as
\begin{equation}
\chi(z) = \langle\langle J; J\rangle\rangle_{z} = i \int_{0}^{\infty} e^{izt} \langle \left[ J(t), J \right] \rangle,
\label{correlator}
\end{equation}
where $J = \sum_{\textbf{k}} e v(\textbf{k})c_{\textbf{k},\sigma}^{\dagger}c_{\textbf{k},\sigma} $ is the current density and $v(\textbf{k})$ is the velocity dispersion. Here $\left[ J(t), J \right]$ denotes the commutator, $\langle ... \rangle$ denotes the ensemble average at temperature $T$ and $\langle\langle ... \rangle\rangle$ denotes the Laplace transform of the ensemble average.

According to the G\"otze and W\"olfle approach\cite{gotze_72}, the memory function is defined as
\begin{equation}
M(z)=z\frac{\chi(z)}{\chi_{0}-\chi(z)} \hspace*{1cm} \textnormal{or} \hspace*{1cm} \chi(z)=\chi_{0}\frac{M(z)}{z+M(z)},
\label{cor}
\end{equation}
where $\chi_{0}$ corresponds to the static limit of correlation function (i.e.  $\chi_{0}=N_{e}/m$)\cite{gotze_72}. Using this, the conductivity from eqn.(\ref{cond}) in terms of memory function can be written as
\begin{equation}
\sigma(z)=\frac{i}{4\pi} \frac{\omega_{p}^{2}}{z+M(z)}.
\label{conddi}
\end{equation}
In ref.\cite{gotze_72}, an expansion for $M(z)=\frac{z\chi(z)}{\chi_{0}} \left(1+\frac{\chi(z)}{\chi_{0}}-...\right)$ is used. Basis of this assumption is the smallness of electron-phonon interaction energy as compared to the Fermi energy of free electrons. Using this expansion and on keeping the leading order term, the memory function $M(z)$ can be written as
\begin{equation}
M(z)=z\frac{\chi(z)}{\chi_{0}} = z\frac{\langle\langle J; J\rangle\rangle_{z}}{\chi_{0}}.
\label{memory66}
\end{equation} 
To compute memory function, we need $\langle\langle J; J\rangle\rangle_{z}$ which by using equation of motion is
\begin{equation}
z\langle\langle J; J\rangle\rangle_{z} = \langle[J,J]\rangle + \langle\langle[J,H_{\textnormal{ep}}];J\rangle\rangle_{z}.
\end{equation}
As the first term of r.h.s is zero, hence the above expression is equivalent to second term which can be further calculated by applying equation of motion.
\begin{equation}
z\langle\langle[J,H_{\textnormal{ep}}];J\rangle\rangle_{z} = \langle[[J,H_{\textnormal{ep}}],J]\rangle - \langle\langle[J,H_{\textnormal{ep}}];[J,H_{\textnormal{ep}}]\rangle\rangle_{z}.\\
\end{equation}
For $z=0$, $\langle[[J,H_{\textnormal{ep}}],J]\rangle=\langle\langle[J,H_{\textnormal{ep}}];[J,H_{\textnormal{ep}}]\rangle\rangle_{z=0}$. Thus, the memory function $M(z)$ becomes
\begin{equation}
M(z)= \frac{\phi(0) - \phi(z)}{z\chi_{0}}.
\label{memory}
\end{equation}
Here $\phi(z)$ (called as correlation function) is defined as
\begin{equation}
\phi(z)=\left\langle \left\langle \left[J,H_{\textnormal{ep}} \right]; \left[J,H_{\textnormal{ep}} \right] \right\rangle \right\rangle_{z}.
\label{gen}
\end{equation}
For the present case of electron-phonon interaction, the correlation function from eqns.(\ref{eb}) and (\ref{gen}) is
\begin{eqnarray} \nonumber
\phi(z)=\sum_{\textbf{k},\textbf{k}'} \sum_{\textbf{p}, \textbf{p}'} \sum_{\sigma, \sigma^{'}} \left[ v_{1}(\textbf{k}) - v_{1}(\textbf{k}')\right]\left[ v_{1}(\textbf{p}) - v_{1}(\textbf{p}')\right] \\ \nonumber
\times \left\langle \left\langle D(\textbf{k}-\textbf{k}')c_{\textbf{k}, \sigma}^{\dagger} c_{\textbf{k}^{\prime}, \sigma} b_{\textbf{k}-\textbf{k}'}-D^{*}(\textbf{k}-\textbf{k}')b^{\dagger}_{\textbf{k}-\textbf{k}'}c_{\textbf{k}', \sigma}^{\dagger} c_{\textbf{k}, \sigma}; \right. \right.\\ \nonumber 
\left. \left.D(\textbf{p}-\textbf{p}')c_{\textbf{p}, \sigma^{'}}^{\dagger} c_{\textbf{p}^{\prime}, \sigma^{'}} b_{\textbf{p}-\textbf{p}'}-D^{*}(\textbf{p}-\textbf{p}')b^{\dagger}_{\textbf{p}-\textbf{p}'}c_{\textbf{p}', \sigma^{'}}^{\dagger} c_{\textbf{p}, \sigma^{'}} \right\rangle \right\rangle \\
\end{eqnarray}
\begin{eqnarray} \nonumber
\phi(z) =-\sum_{\textbf{k},\textbf{k}'} \sum_{\textbf{p}, \textbf{p}'} \sum_{\sigma, \sigma^{'}} \left[ v_{1}(\textbf{k}) - v_{1}(\textbf{k}')\right]\left[ v_{1}(\textbf{p}) - v_{1}(\textbf{p}')\right] \\ \nonumber
\times \left[D(\textbf{k}-\textbf{k}')D^{*}(\textbf{p}-\textbf{p}') \right] \\ \nonumber
\times \left( \left\langle \left\langle c^{\dagger}_{\textbf{k}, \sigma} c_{\textbf{k}', \sigma} b_{\textbf{k}-\textbf{k}'}; b^{\dagger}_{\textbf{p}-\textbf{p}'} c^{\dagger}_{\textbf{p}', \sigma^{'}} c_{\textbf{p}, \sigma^{'}}  \right\rangle \right\rangle \right. \\ \nonumber
\left.+ \left\langle \left\langle c_{\textbf{k}, \sigma} c^{\dagger}_{\textbf{k}', \sigma} b^{\dagger}_{\textbf{k}-\textbf{k}'};b_{\textbf{p}-\textbf{p}'} c_{\textbf{p}', \sigma^{'}} c^{\dagger}_{\textbf{p}, \sigma^{'}} \right\rangle \right\rangle \right).\\
\label{pz}
\end{eqnarray}
To evaluate the $\phi(z)$, we need to calculate \\
$\left\langle \left\langle c^{\dagger}_{\textbf{k}, \sigma} c_{\textbf{k}', \sigma} b_{\textbf{k}-\textbf{k}'}; b^{\dagger}_{\textbf{p}-\textbf{p}'}c^{\dagger}_{\textbf{p}', \sigma^{'}} c_{\textbf{p}, \sigma^{'}}  \right\rangle \right\rangle $ which can be calculated as (using definition \ref{correlator})
\begin{eqnarray} \nonumber
\left\langle \left\langle c^{\dagger}_{\textbf{k}, \sigma} c_{\textbf{k}', \sigma} b_{\textbf{k}-\textbf{k}'}; b^{\dagger}_{\textbf{p}-\textbf{p}'}c^{\dagger}_{\textbf{p}', \sigma^{'}} c_{\textbf{p}, \sigma^{'}}  \right\rangle \right\rangle = \\ \nonumber
i \int_{0}^{\infty} dt e^{izt} \langle[c^{\dagger}_{\textbf{k}, \sigma}(t) c_{\textbf{k}', \sigma}(t) b_{\textbf{k}-\textbf{k}'}(t) \\  ;b^{\dagger}_{\textbf{p}-\textbf{p}'}c^{\dagger}_{\textbf{p}', \sigma^{'}} c_{\textbf{p}, \sigma^{'}}]\rangle.
\end{eqnarray}
Using $c_{\textbf{k}, \sigma}(t)=c_{\textbf{k}, \sigma}e^{-i\epsilon_{\textbf{k}}t}$ and performing the integration over time and ensemble average, we have
\begin{eqnarray} \nonumber
\left\langle \left\langle c^{\dagger}_{\textbf{k}, \sigma} c_{\textbf{k}', \sigma} b_{\textbf{k}-\textbf{k}'};b^{\dagger}_{\textbf{p}-\textbf{p}'} c^{\dagger}_{\textbf{p}', \sigma^{'}} c_{\textbf{p}, \sigma^{'}} \right\rangle \right\rangle = \\
- \frac{\left[f(1-f')(1+n) - f'(1-f)n\right] \delta_{\textbf{k},\textbf{p}} \delta_{\textbf{k}',\textbf{p}'}\delta_{\sigma, \sigma^{'}}}{z-\epsilon_{\textbf{k}'}+\epsilon_{\textbf{k}} - \omega_{\textbf{k}-\textbf{k}'}}.
\end{eqnarray}
Here $f\equiv f(\epsilon_{\textbf{k}})=\left(e^{\beta\epsilon_{\textbf{k}}}+1\right)^{-1}$ and $n\equiv n(\omega_{\textbf{k}-\textbf{k}^{\prime}})= \left(e^{\beta\omega_{\textbf{k}-\textbf{k}'}}-1\right)^{-1}$ represent the Fermi and Bose distribution functions and $\beta$ corresponds to inverse of temperature. Inserting this equation in eqn.(\ref{pz}) and hence in eqn.(\ref{memory}) and then by taking the limit $z \rightarrow \omega+i\eta$, $\eta \rightarrow 0^{+}$, the imaginary part of the memory function can be expressed as
\begin{eqnarray} \nonumber
M''(\omega, T)&=&\frac{2\pi}{3} \frac{1}{mN_{e}} \sum_{\textbf{k},\textbf{k}'} \vert D(\textbf{k}-\textbf{k}') \vert ^{2} (\textbf{k}-\textbf{k}')^{2} f'(1-f)n \\ \nonumber
&& \left[ \frac{e^{\beta\omega}-1}{\omega} \delta(\epsilon_{\textbf{k}}-\epsilon_{\textbf{k}'}-\omega_{\textbf{k}-\textbf{k}'}+ \omega)\right. \\
&& \left. +(\textnormal{terms with} \,  \omega \rightarrow -\omega) \right].
\label{imagmem}
\end{eqnarray}
Convert the summations over $\textbf{k}$ and $\textbf{k}'$ into integrations and assuming that $\textbf{k}$ is pointing along the z-direction and $\textbf{k}^{'}$ subtends an angle $\theta$ with it (at the end \textbf{k} integration over all directions and magnitudes is to be performed). Insert an integral $\int dq \delta(q-\vert \textbf{k}-\textbf{k}' \vert)$ over $q$ to stratify the calculation as given below. Thus the eqn.(\ref{imagmem}) becomes
\begin{eqnarray} \nonumber
M''(\omega, T) &=&  \frac{2}{3} \pi \frac{2 N^{2}}{(2\pi)^{4} m N_{e}} \int_{0}^{\infty} dq q^{2} \vert D(q)\vert ^{2}\int_{0}^{\infty} dk k^{2} \\ \nonumber
&&  \int_{0}^{\infty} dk' k'^{2} \int_{0}^{\pi} d\theta \sin\theta \delta(q-\vert \textbf{k}-\textbf{k}' \vert) \\ \nonumber
&& f'(1-f)n \left[ \frac{e^{\beta\omega}-1}{\omega} \delta(\epsilon_{\textbf{k}}-\epsilon_{\textbf{k}'}-\omega_{\textbf{k}-\textbf{k}'}+ \omega) \right.\\
&& \left. +(\textnormal{terms with} \,  \omega \rightarrow -\omega) \right].
\end{eqnarray}
Here due to the presence of Fermi factors $f'(1-f)$ the integrand has finite value only around the Fermi surface and vanishes outside the strip of width $2/\beta$ $(\omega << \epsilon_{F})$. Thus $\textbf{k}$ and $\textbf{k}'$ can be approximately replaced by $\textbf{k}_{F}$. With this the $\theta$ integral can be simplified as $\int_{0}^{\pi} d\theta \sin\theta \delta(q-\sqrt{2}k_{F}\sqrt{1-\cos\theta})$ which will yield the result $\frac{q}{k_{F}^{2}}$. Using this and converting $\textbf{k}$ integrals into energy integrals, the above equation reduces to
\begin{eqnarray} \nonumber
M''(\omega, T) &=& \frac{4}{3} \frac{N^{2} m^{2}\epsilon_{F}}{(2\pi)^{3} N_{e}k_{F}^{2}} \int_{0}^{q_{D}} dq q^{3} \vert D(q) \vert ^{2} \\ \nonumber
&& \int_{-\infty}^{\infty} d\epsilon  \frac{n}{e^{-\beta(\epsilon-\epsilon_{F})}+1}  \\ \nonumber
&& \left[\frac{1}{e^{\beta(\epsilon-\epsilon_{F}+\omega-\omega_{q})}+1} \frac{e^{\beta\omega}-1}{\omega} \right.\\
&& \left. + (\textnormal{terms with} \, \omega \rightarrow -\omega)\right].
\label{mem}
\end{eqnarray}
This is an expression for the imaginary part of memory function as deduced by G\"otze-W\"olfle\cite{gotze_72}. It can be simplified by using electron-phonon matrix element for acoustic phonons which is defined as\cite{ziman_60}
\begin{equation}
D(\textbf{q}) = \left(\frac{1}{2m_{i}N\omega_{q}}\right)^{-1/2} q C(q); \hspace{0.5cm} \omega_{q} = c_{s}q,
\label{disp}
\end{equation}
where $C(q)$ is the slowly varying function of $q$, $m_{i}$ is the ion mass, $N$ is the total number of unit cells and $c_{s}$ is the sound velocity. To analyze the eqn.(\ref{mem}), various limiting cases using eqn.(\ref{disp}) were discussed in ref.\cite{gotze_72}. 
\section{Memory Function with Gapped Density of states}
\label{sec:2}
In this section we go beyond the assumption of constant electronic density of states and we consider a system with a gap around the Fermi surface. In this case, density of states is zero in energy region ($-\Delta,\Delta$). Thus the energy integration in eqn.(\ref{mem}) has to be
\begin{eqnarray} \nonumber
I &=& \int_{-\infty}^{\epsilon_{F}-\Delta} d\epsilon \frac{e^{\beta(\epsilon-\epsilon_{F})}}{e^{\beta(\epsilon-\epsilon_{F})}+1}  \frac{1}{e^{\beta(\epsilon-\epsilon_{F}+\omega-\omega_{q})}+1} \\
&+& \int_{\epsilon_{F}+\Delta}^{\infty} d\epsilon \frac{e^{\beta(\epsilon-\epsilon_{F})}}{e^{\beta(\epsilon-\epsilon_{F})}+1}  \frac{1}{e^{\beta(\epsilon-\epsilon_{F}+\omega-\omega_{q})}+1}.
\end{eqnarray} 
After simplification we have
\begin{eqnarray} \nonumber
I &=&\frac{1}{\beta}\frac{1}{e^{\beta(\omega-\omega_{q})}-1} \\
&& \log \left(\frac{(1+e^{\beta(-\Delta+\omega-\omega_{q})})(1+e^{\beta\Delta})e^{\beta(\omega-\omega_{q})}}{(1+e^{\beta(\Delta+\omega-\omega_{q})})(1+e^{-\beta\Delta})} \right).
\end{eqnarray}
Using this, the imaginary part of memory function can be written as
\begin{eqnarray}\nonumber
M''(\omega, T) &=& \frac{\pi^{3} N^{2} \rho_{F}^{2}}{4mk_{F}^{5}} \int_{0}^{q_{D}} dq q^{3} \vert D(q) \vert ^{2}\frac{1}{\beta}n\\ \nonumber
&& \left[ \frac{e^{\beta\omega}-1}{\omega} \frac{1}{e^{\beta(\omega-\omega_{q})}-1} \right. \\ \nonumber
&&\left. \times \log \left[ \left( \frac{1+e^{\beta\Delta}}{1+e^{-\beta\Delta}} \right) \left(\frac{1+e^{-\beta(\Delta-\omega+\omega_{q})}}{e^{\beta \Delta}+e^{\beta(\omega_{q}-\omega)}} \right)\right] \right.\\
&& \left. + (\textnormal{terms with} \,  \omega \rightarrow -\omega) \right]. 
\label{greater}
\end{eqnarray}
This is the desired expression for the frequency and temperature dependent imaginary part of memory function. For $\Delta=0$ and using phonon matrix element (eqn.(\ref{disp})), this expression reduces to  the expression (refer eqn.(54(a))) given in original GW work\cite{gotze_72}, as it should. In actual practise (i.e. for an arbitrary form of gap around the Fermi surface), the general expression of the imaginary part of memory function is complicated and is difficult to proceed analytically. A general formulae is given in appendix(\ref{Appendix}). Thus for the simplicity of calculation, we have discussed the specific system in this article. Further to write $M''(\omega,T)$ in compact form, change the variable $\omega_{q}$ to $\Omega$ in above equation which can be rewritten as
\begin{eqnarray} \nonumber
M''(\omega, T) &=& \frac{2\pi}{\omega} \int_{0}^{\omega_{D}} d\Omega \alpha^{2}F(\Omega) \frac{1}{\beta}\left[\frac{e^{\beta\omega}-1}{e^{\beta(\omega-\Omega)}-1} \right. \\ \nonumber
&& \left.  \frac{1}{e^{\beta\Omega}-1}\log \left(\frac{1+e^{-\beta(\Delta-\omega+\Omega)}}{1+e^{-\beta(\Delta+\omega-\Omega)}}\right) \right. \\
&& \left. - (\textnormal{terms with} \, \omega \rightarrow -\omega)\right],
\label{memome}
\end{eqnarray} 
where $\alpha^{2}F(\Omega)$ is defined as
\begin{equation}
\alpha^{2}F(\Omega)= \frac{\pi^{2} N^{2} \rho_{F}^{2}}{8mk_{F}^{5}c_{s}^{4}} \Omega^{3} \vert D(\Omega) \vert ^{2}.
\end{equation}
This is known as phonon spectral function\cite{allen_71}. In the case of cuprates, it is replaced by $I^{2}\chi(\Omega)$ which represents the boson spectral function\cite{sharapov_05}. This form is same as given by Allen \cite{allen_71} $\left(\alpha^{2}F(\Omega)= \frac{N(0)}{4v_F^2}\langle \langle \mid M_{\textbf{k}\textbf{k}'}\mid ^2 (v(\textbf{k})-v(\textbf{k}^{\prime}))^2\right.$ \\
$\left.\delta(\hbar\Omega_{\textbf{Q}} -\hbar\Omega)\rangle\rangle\right)$.

Equation (\ref{memome}) is our main result. To discuss it in various temperature and frequency regimes, we use the phonon matrix element eqn.(\ref{disp})  and calculate $M''(\omega, T)$ in next subsections.
\subsection{DC memory function}
\label{sec:3a}
In the zero frequency limit and assuming $C(q)$ as a constant i.e. $C(q)=1/\rho_{F}$ \cite{ziman_60}, the imaginary part of the memory function (eqn.(\ref{greater})) becomes
\begin{eqnarray} \nonumber
M''(0, T)&=& \frac{1}{8} \pi^{3} \frac{N}{m m_{i}k_{F}^{5}}\int_{0}^{q_{D}} dq q^{5} \frac{1}{(e^{\beta\omega_{q}}-1)(e^{-\beta\omega_{q}}-1)} \\ \nonumber
&&\frac{1}{\omega_{q}}\log \left[ \frac{1+e^{\beta\Delta}}{1+e^{-\beta\Delta}} \frac{1+e^{-\beta(\Delta + \omega_{q})}}{e^{\beta \Delta}+e^{\beta\omega_{q}}} \right]. \\
\label{dcc}
\end{eqnarray}
Now consider the case of $T >> \omega_{D}, \Delta$, the above equation reduces to
\begin{eqnarray}\nonumber
M''(0, T) &=& \frac{1}{8} \pi^{3} \frac{N}{m m_{i}k_{F}^{5}} \int_{0}^{q_{D}} dq q^{5} \frac{1}{\omega_{q}} \frac{-1+\beta\omega_{q}}{(\beta\omega_{q})^{2}} \\
&& \times \log \left[\frac{2-\beta\Delta-\beta\omega_{q}}{2-\beta\Delta+\beta\omega_{q}} \right].
\end{eqnarray}
On substituting $x=\frac{q\Theta_{D}}{q_{D}T}$ (i.e. $\beta\omega_{q}=x$) where $\Theta_{D}$ is the Debye temperature, the dc memory function reduces to
\begin{eqnarray} \nonumber
M''(0,T)&=& \frac{1}{8} \pi^{3} \frac{N}{m m_{i}k_{F}^{5} \Theta_{D}} q_{D}^6 \left(\frac{T}{\Theta_{D}}\right)^{5} \\ \nonumber
&&\int_{0}^{\beta\Theta_{D}} dx x^{2}(x-1) \log\left[ \frac{2-\beta \Delta -x}{2-\beta\Delta +x} \right]. \\
\label{low1}
\end{eqnarray}
This expression under case $T >> \omega_{D}, \Delta$ is equivalent to
\begin{equation}
M''(0,T) \simeq A \left\lbrace \frac{T}{\Theta_{D}} + \frac{\Delta}{\Theta_{D}} +\frac{1}{T} \left( \frac{\Delta^{2}}{8\Theta_{D}} + \frac{8\Delta}{5}-\frac{\Theta_{D}}{6}\right)...\right\rbrace.
\end{equation}
where $A$ refers for constant numerical factor.
Similarly for $T<< \omega_{D}, \Delta$, the eqn.(\ref{dcc}) becomes
\begin{eqnarray} \nonumber
M''(0, T)&=& -\frac{1}{8} \pi^{3} \frac{N}{m m_{i}k_{F}^{5} \Theta_{D}} q_{D}^6 \left(\frac{T}{\Theta_{D}}\right)^{5} \\
&&\int_{0}^{\beta\Theta_{D}} dx x^{4}e^{-x} \log\left[ \frac{e^{\beta \Delta}+e^{-x}}{e^{\beta\Delta}+e^{x}} \right].
\label{high1}
\end{eqnarray}
This expression can also be simplified as
\begin{equation}
M''(0,T) \simeq A e^{-\beta \Delta} \left\lbrace \frac{1}{5} -  \frac{3}{4} \left( \frac{T}{ \Theta_{D}} \right)^{5} ... \right\rbrace.
\label{highlimit}
\end{equation}
Substituting the equation (\ref{high1}) in eqn.(\ref{cor}) and hence in eqn.(\ref{cond}), leads to the expression of dc conductivity for the electron-phonon interaction. Here if we insert gap $\Delta=0$ in the eqn.(\ref{dcc}), we obtain eqn.(54(b)) as given in ref.\cite{gotze_72}, as expected.
\subsection{AC memory function}
\label{sec:3b}
We proceed again with eqn.(\ref{greater}) to study frequency dependent behaviour of memory function in different regimes. In the high frequency regime i.e. for $\omega >> \omega_{D}$ and using same approximation ($C(q)=1/\rho_{F}$) as considered for the dc case, the imaginary part of memory function becomes
\begin{eqnarray} \nonumber
M''(\omega, T) &=& \frac{1}{8} \pi^{3} \frac{N}{mm_{i}k_{F}^{5}}  \int_{0}^{q_{D}} dq q^{5} \frac{1}{\beta \omega_{q}}\frac{n}{\omega} \\ \nonumber
&&\left[ \log\left(\frac{1+e^{-\beta(\Delta-\omega)}}{1+e^{-\beta(\Delta+\omega)}}\right) \right. \\ 
&& \left.- e^{\beta\omega_{q}} \log\left(\frac{1+e^{-\beta(\Delta+\omega)}}{1+e^{-\beta(\Delta - \omega)}} \right) \right].
\end{eqnarray}
When the gap is smaller than the $\vert \omega-\omega_{D}\vert$ i.e. $\Delta < \vert \omega-\omega_{D} \vert$, the above equation reduces to \footnote{In the opposite case $\vert \omega - \omega_{D} \vert < \Delta$, eqn.(\ref{greater}) leads to vanishing scattering rate.}
\begin{eqnarray}\nonumber
M''(\omega, T) &=& \frac{1}{8} \pi^{3} \frac{N}{m m_{i}k_{F}^{5} \Theta_{D}} q_{D}^6 \left(\frac{T}{\Theta_{D}}\right)^{5} \\ 
&& \int_{0}^{\beta\Theta_{D}} dx x^{4} \coth\left(\frac{x}{2}\right).
\label{acmem}
\end{eqnarray}
From this we identify that at high temperature, the imaginary part of memory function becomes temperature and frequency independent. This means the saturation behavio-ur of $M''(\omega,T)$ for $\omega >> \omega_{D}$. The reason is that under this condition, the integral approaches to $(\Theta_{D}/T)^{5}$ and it cancels with prefactor $(T/\Theta_{D})^{5}$ in eqn.(\ref{acmem}). At low temperature, it varies linearly with temperature as the integral approaches to $(\Theta_{D}/T)^{4}$.

In the next section we compare our findings (eqn.(\ref{memome})) with the Sharapov-Carbotte\cite{sharapov_05}.
\section{Comparison with Sharapov-Carbotte results}
\label{sec:3}
Sharapov and Carbotte has deduced a relation for the generalized Drude scattering rate\cite{sharapov_05} taking electron-boson interaction and non constant electronic density of states. Using Kubo formula\cite{kubo_57} and calculating the self energy under certain assumptions (as discussed below), they derived the following expression
\begin{eqnarray} \nonumber
\frac{1}{\tau(\omega, T)} &=& \frac{\pi}{\omega} \int_{0}^{\infty} d\Omega I^{2}\chi(\Omega) \int_{-\infty}^{\infty} d\omega^{\prime} \\ \nonumber
&&\left[ \frac{\tilde{N}(\omega^{\prime}-\Omega)}{N(0)} + \frac{\tilde{N}(-\omega^{\prime}+\Omega)}{N(0)} \right] \\ \nonumber
&& \times \left[n(\Omega)+ f(\Omega-\omega^{\prime}) \right]\\
&& \times \left[f(\omega^{\prime}-\omega)-f(\omega^{\prime}+\omega) \right],
\label{shara}
\end{eqnarray}
where $I^{2}\chi(\omega)$ is the boson spectral function and $\tilde{N}(\omega)$ is the quasiparticle electronic density of states and $N(0)$ is for normalization. In deriving the above formula, the following assumptions were made: (1) vertex corrections were neglected, (2) energy independent character of plasma frequency in the vicinity of Fermi level and (3) $\vert \Sigma(\epsilon+\omega) - \Sigma^{*}(\epsilon) \vert << \omega$ where $\Sigma(\epsilon)$ is the electronic self energy.

\begin{figure}[htb]
\centering
\hspace{-0cm}
\subfigure[noonleline][]
{\label{schm1}\includegraphics[height=30mm,width=40mm]{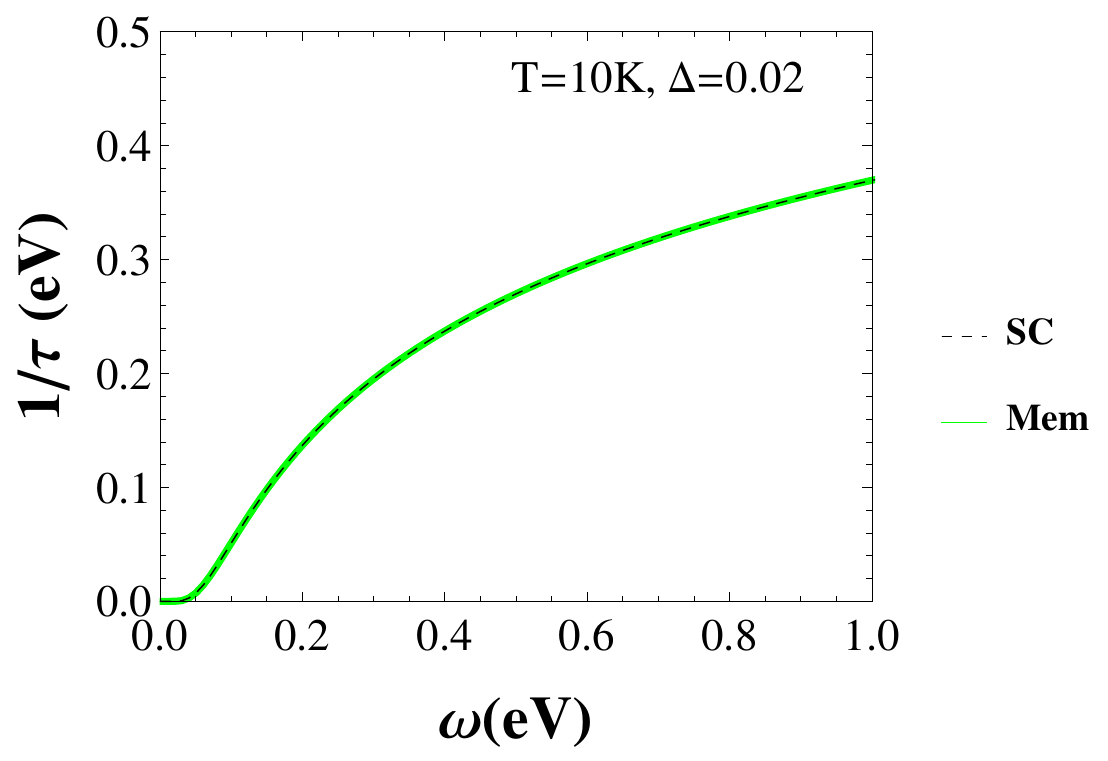}}
\hspace{-0.2cm}
\subfigure[noonleline][]
{\label{schm}\includegraphics[height=30mm,width=40mm]{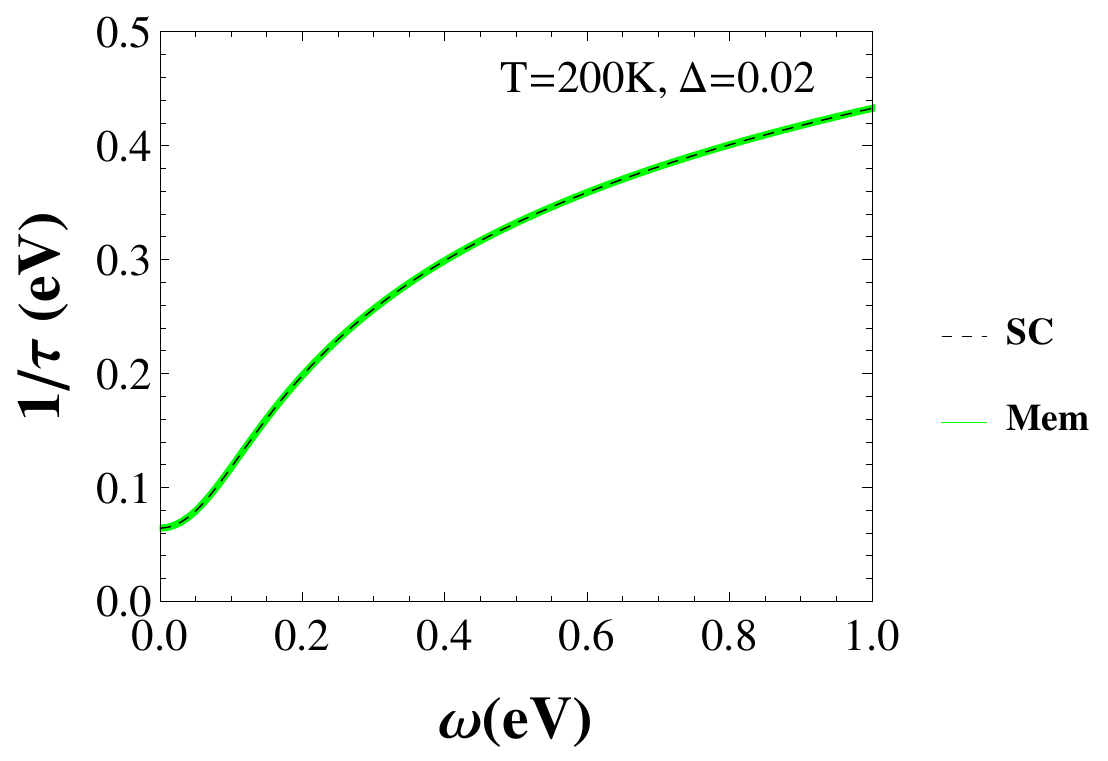}}
\hspace{0cm}
\subfigure[noonleline][]
{\label{schm2}\includegraphics[height=30mm,width=40mm]{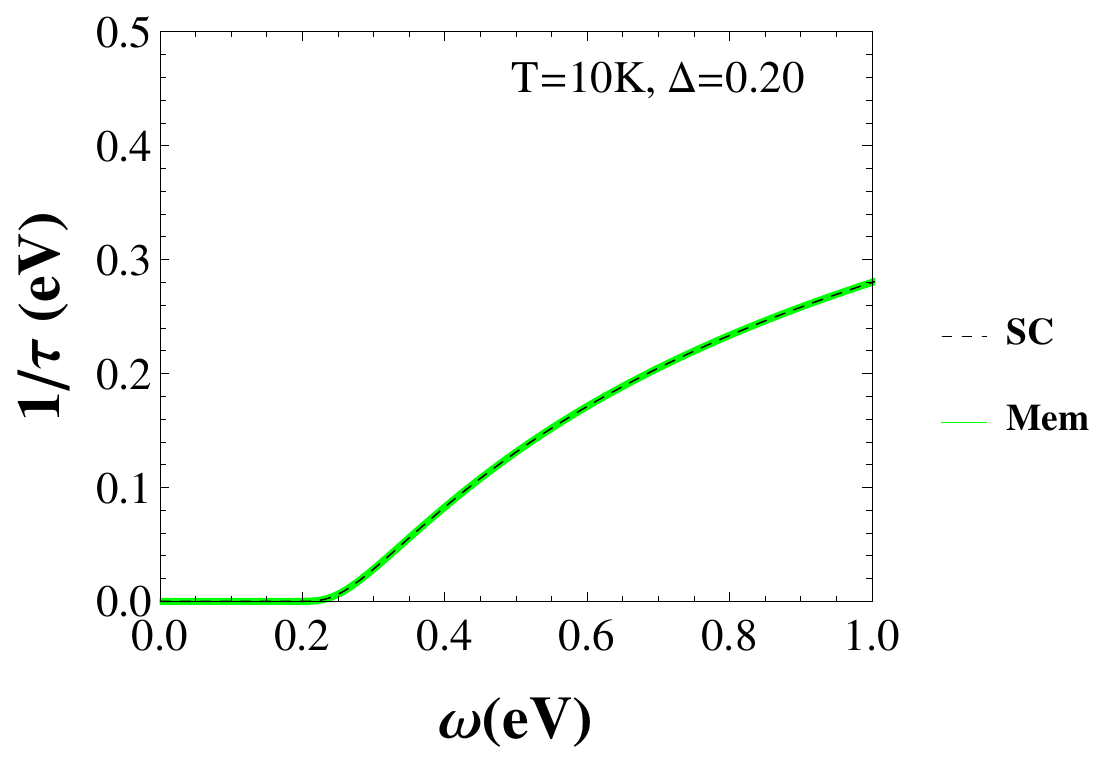}}
\hspace{-0.2cm}
\subfigure[noonleline][]
{\label{schm3}\includegraphics[height=30mm,width=40mm]{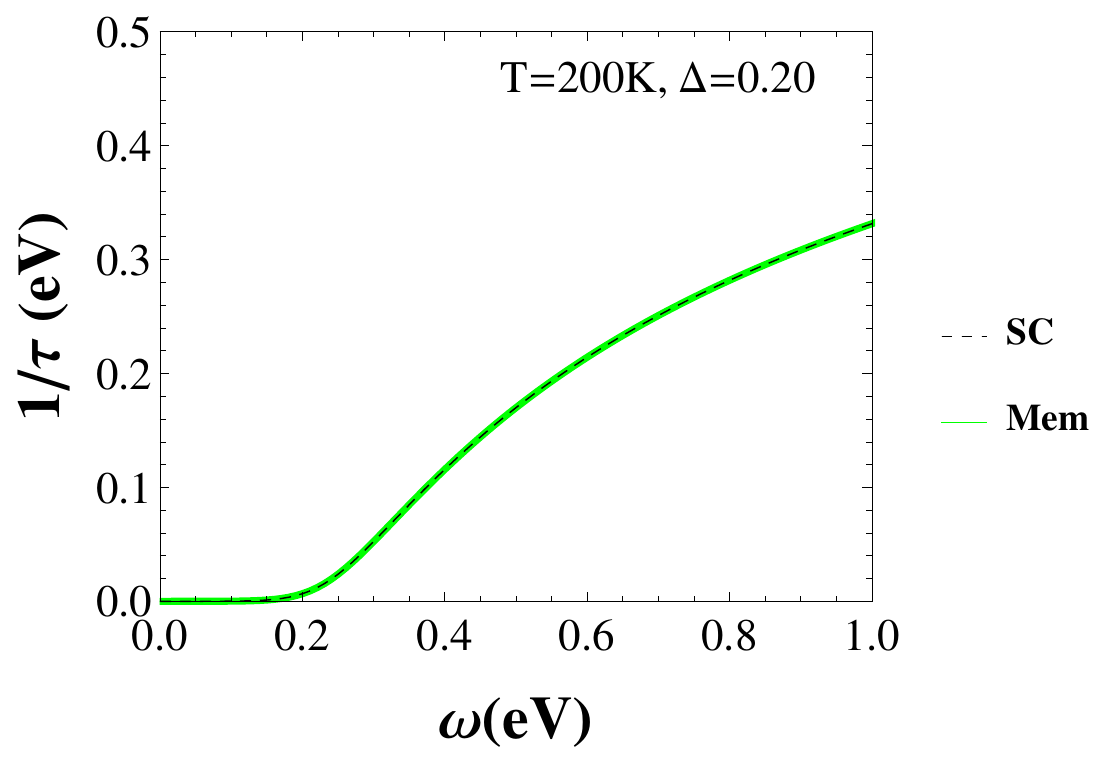}}
\caption{Comparison plot of scattering rate $1/\tau(\omega,T)$(=$M''(\omega,T)$) calculated using Memory function approach (Mem, solid, green) and Sharapov-Carbotte approach (SC, dashed, black) at temperature T$=10$K and $200$K and at gap $\Delta=0.02$eV and $0.20$eV. The agreement is excellent.}
\label{temp}
\end{figure}
\begin{figure}[htb]
\centering
\hspace*{-0cm}
\subfigure[noonleline][]
{\label{schm4}\includegraphics[height=30mm,width=40mm]{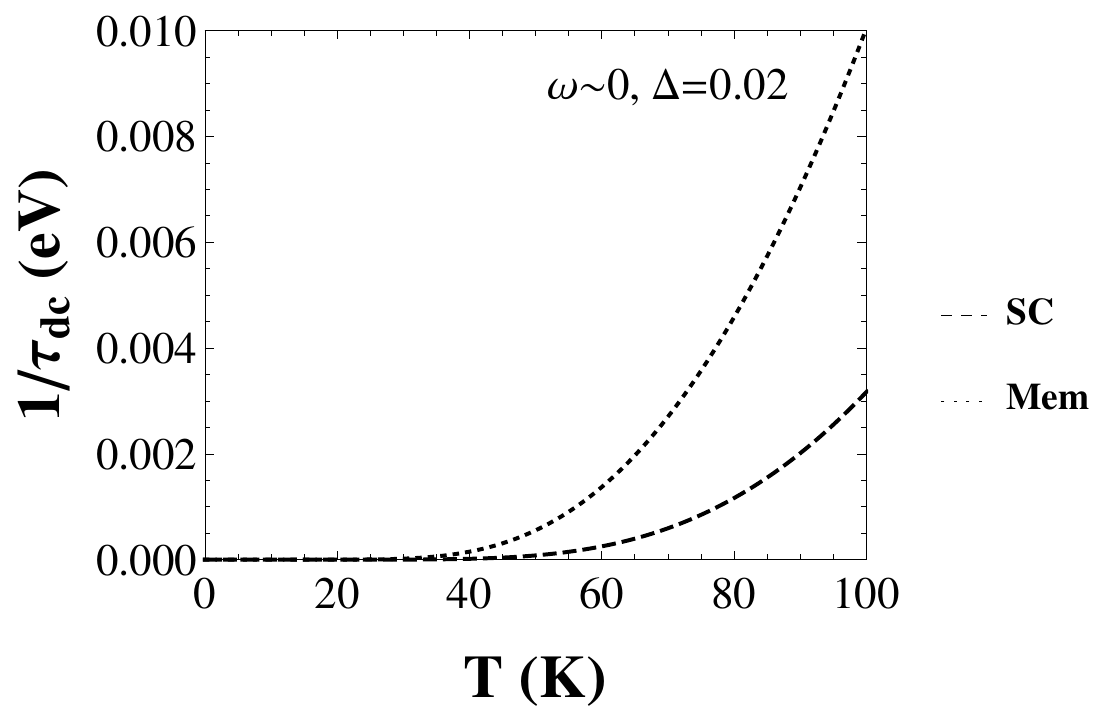}}
\hspace{-0.2mm}
\subfigure[noonleline][]
{\label{schm5}\includegraphics[height=30mm,width=40mm]{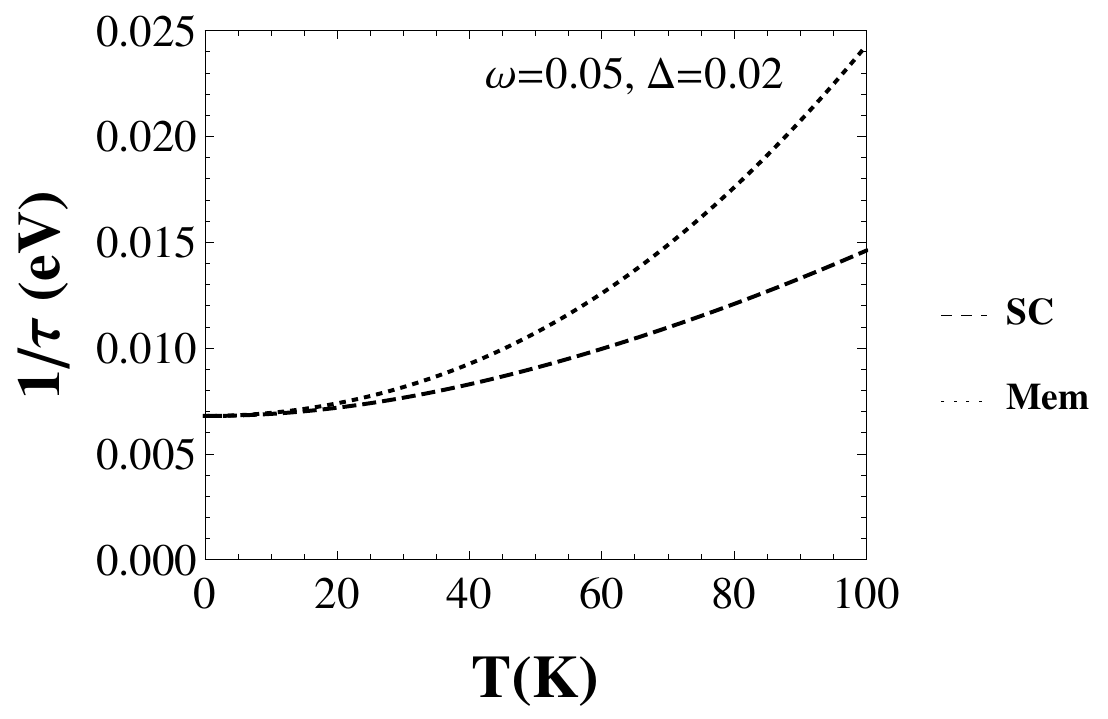}}
\hspace{0.1mm}
\subfigure[noonleline][]
{\label{schm6}\includegraphics[height=30mm,width=40mm]{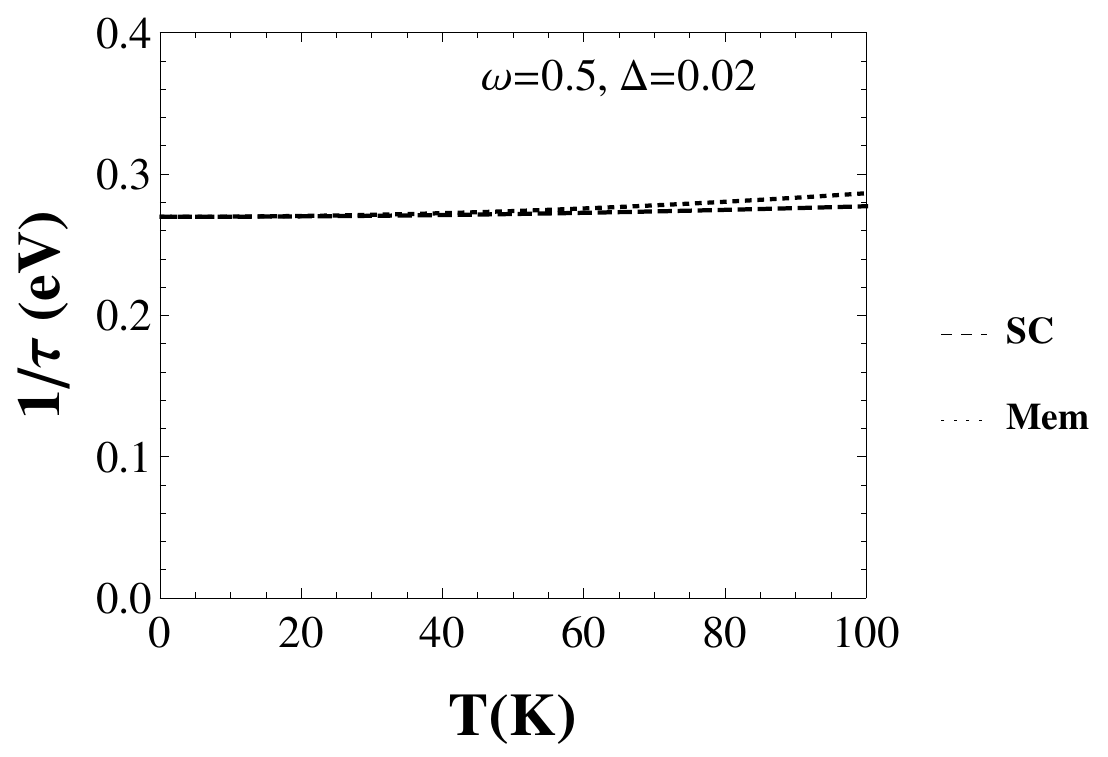}}
\caption{Temperature variation of scattering rate with two different approaches namely memory function (Mem, dotted) and Sharapov-Carbotte (SC, dashed) at gap $0.02$eV. (a) dc case (b) at $\omega = 0.05$eV (c) at $\omega = 0.5$eV.}
\label{tt}
\end{figure}
To compare our approach (eqn.(\ref{memome})) with SC approach (eqn.(\ref{shara})), we have done calculations using models for electronic density of states and the boson spectral function. First in SC approach, for the electronic density of states, we use a square well type model with center at Fermi energy and considered a gap of $2\Delta$ around it. Same gap is taken in our approach (eqn.(\ref{memome})). Second, for the boson spectral function, $I^{2}\chi(\Omega)$, we modelled it as Lorentz-ian of the type $\frac{\Gamma\Omega}{(\Omega-\Omega_{E})^{2}+(\Gamma)^{2}}$ where $\Omega_{E}$ represent the boson peak frequency and $\Gamma$ is the width of the Lorentzian (this form has been used extensively in ref.\cite{hwang_11,hwang_12,bhalla_14}). Thus for comparison, we use the same form of $I^{2}\chi(\Omega)$ in SC approach and our approach. In the whole analysis, we have fixed the value of $\Omega_{E}$ and $\Gamma$ as $0.02$eV and $0.04$eV respectively in both approaches. The value of Debye frequency (the upper limit of phonon frequency integration eqn.(\ref{memome})) is very much high as compared to the Lorentzian width, hence $\omega_{D}$ doesnot give any effect in whole calculation. To compare the results from both the approaches, the frequency dependent scattering rate has been plotted at different temperatures. In fig.(\ref{temp}), we can observe an excellent agreement between both the approaches. As the gap magnitude is increased, the scattering rate shows suppression upto the frequency $\omega \sim \Delta$ as expected (compare figures \ref{schm1} and \ref{schm2}). These results are qualitatively agrees with experimental results\cite{dai_12,lee_05}. \\

In fig.(\ref{tt}), we plot $1/\tau(\omega \rightarrow 0, T)$ as a function of temperature $T$. Here we can observe that the memory function approach yields more magnitude over the SC approach. In fig.(\ref{schm4}) i.e. in zero frequency limit, the ratio $\left\vert \frac{1/\tau_{MF}-1/\tau_{SC}}{/1/\tau_{MF}} \right\vert_{100 K}$, where $1/\tau_{MF}$ and$1/\tau_{SC}$ represents the scattering rate by memory function technique and SC technique respectively, is 0.7 which becomes 0.4 at $\omega = 0.05$eV (as shown in fig.(\ref{schm5})) and at $\omega = 0.5$eV it further reduce to $0.031$ (as shown in fig.(\ref{schm6}). This shows that the difference between scattering rates using memory function approach and SC approach reduces as we go from dc limit to finite frequency limit. Also both approaches explain the Holstein's mechanism at $T=0$K \cite{holstein_54,joyce_70} (as shown in fig. (\ref{schm5}, \ref{schm6})). Thus we notice that there are discrepancies between the two approaches in the d.c. limit. The reasons are discussed in next section.\\

\begin{figure}[htb]
\centering
\hspace{-0cm}
\subfigure[noonleline][]
{\label{dccond1}\includegraphics[height=30mm,width=40mm]{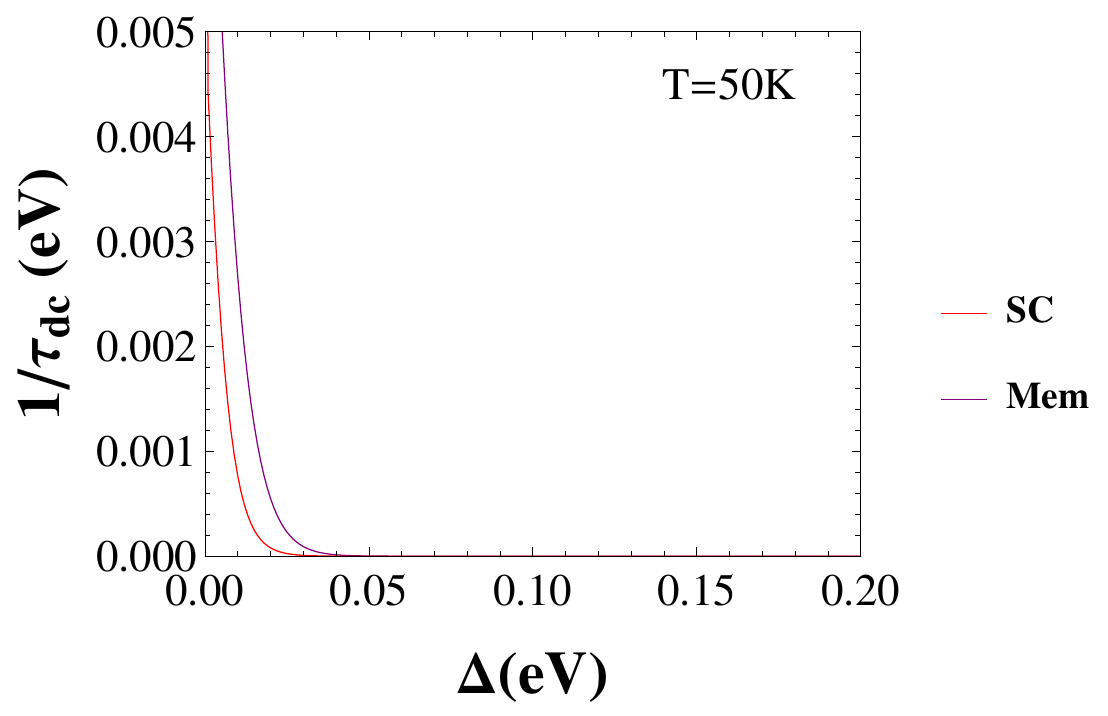}}
\hspace{-0.2cm}
\subfigure[noonleline][]
{\label{dccond2}\includegraphics[height=30mm,width=40mm]{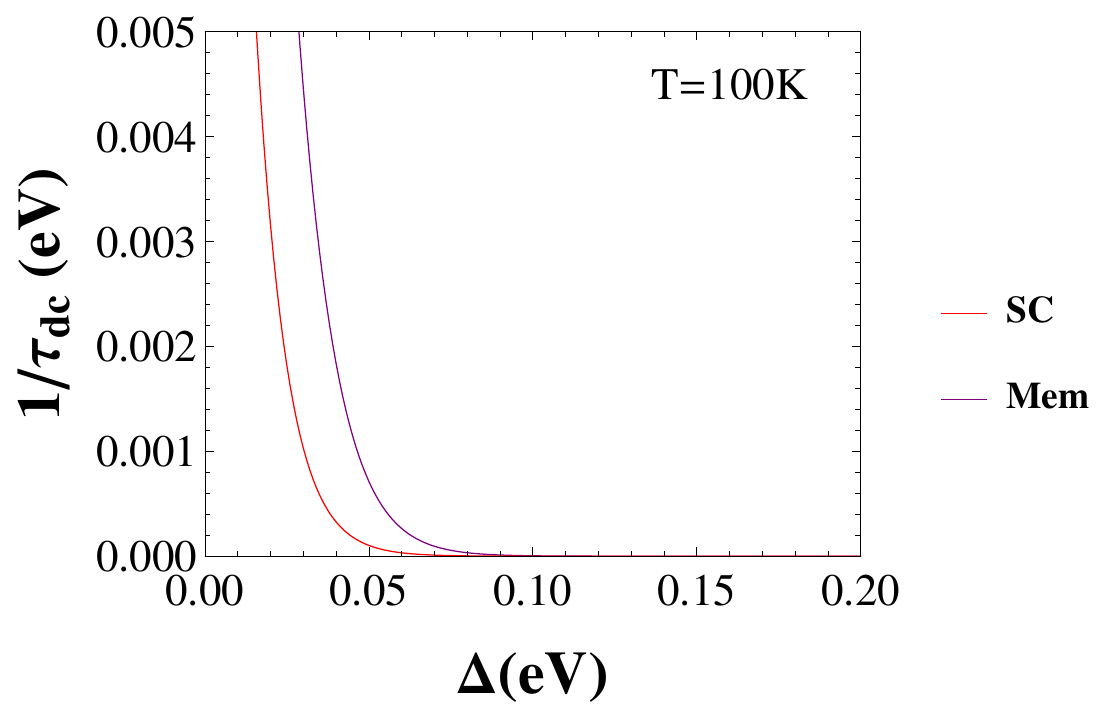}}
\hspace{-0.2cm}
\subfigure[noonleline][]
{\label{dccond4}\includegraphics[height=30mm,width=40mm]{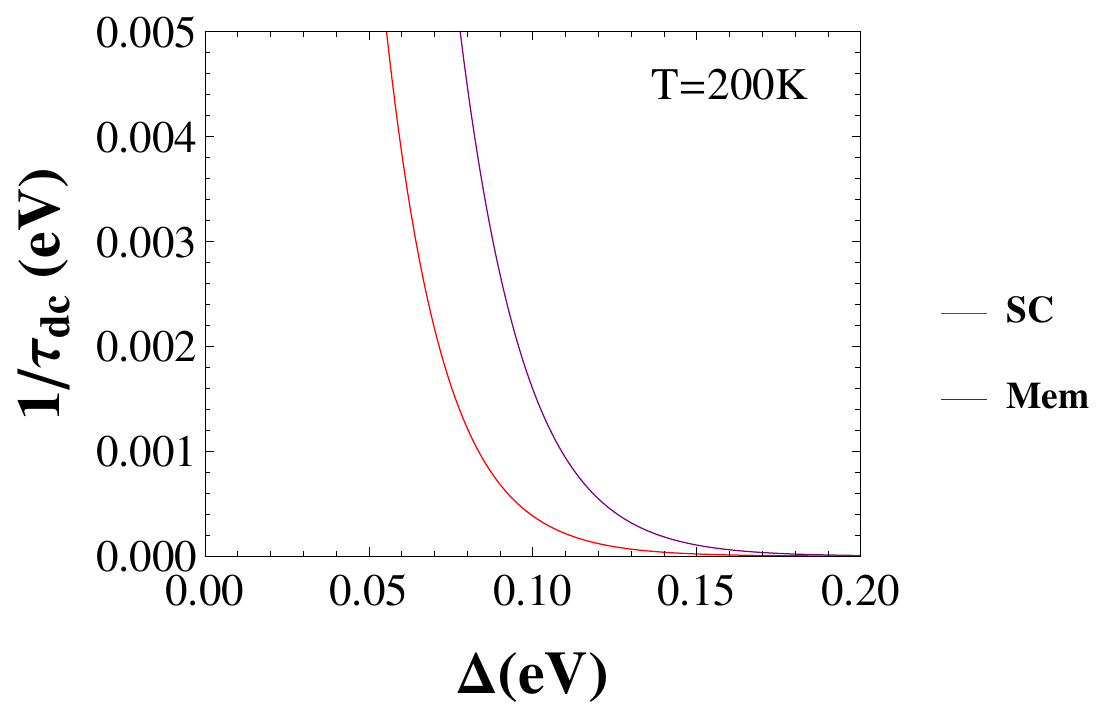}}
\caption{Comparison of dc scattering rate ($1/\tau_{dc}$) as a function of $\Delta$ using Memory function approach (Mem, Purple) and Sharapov-Carbotte approach (SC, Red) at various temperatures 50K, 100K and 200K.}
\label{dcconductivity}
\end{figure}
Next, we have plotted $1/\tau_{dc}$ at different temperatures as a function of $\Delta$ and compare the both approaches. Here we observe that $1/\tau_{dc}$ decreases with the increase of gap energy $\Delta$. Also, we find that the difference between the $1/\tau_{SC}$ and $1/\tau_{MF}$ is not much dependent on $\Delta$, but it does increase with increasing temperature. These discrepencies observed in the dc limit are discussed in the next section.\\
\section{Discussion}
\label{sec:4}
\begin{figure}[htb]
\centering
\hspace{-0cm}
\includegraphics[height=40mm,width=60mm]{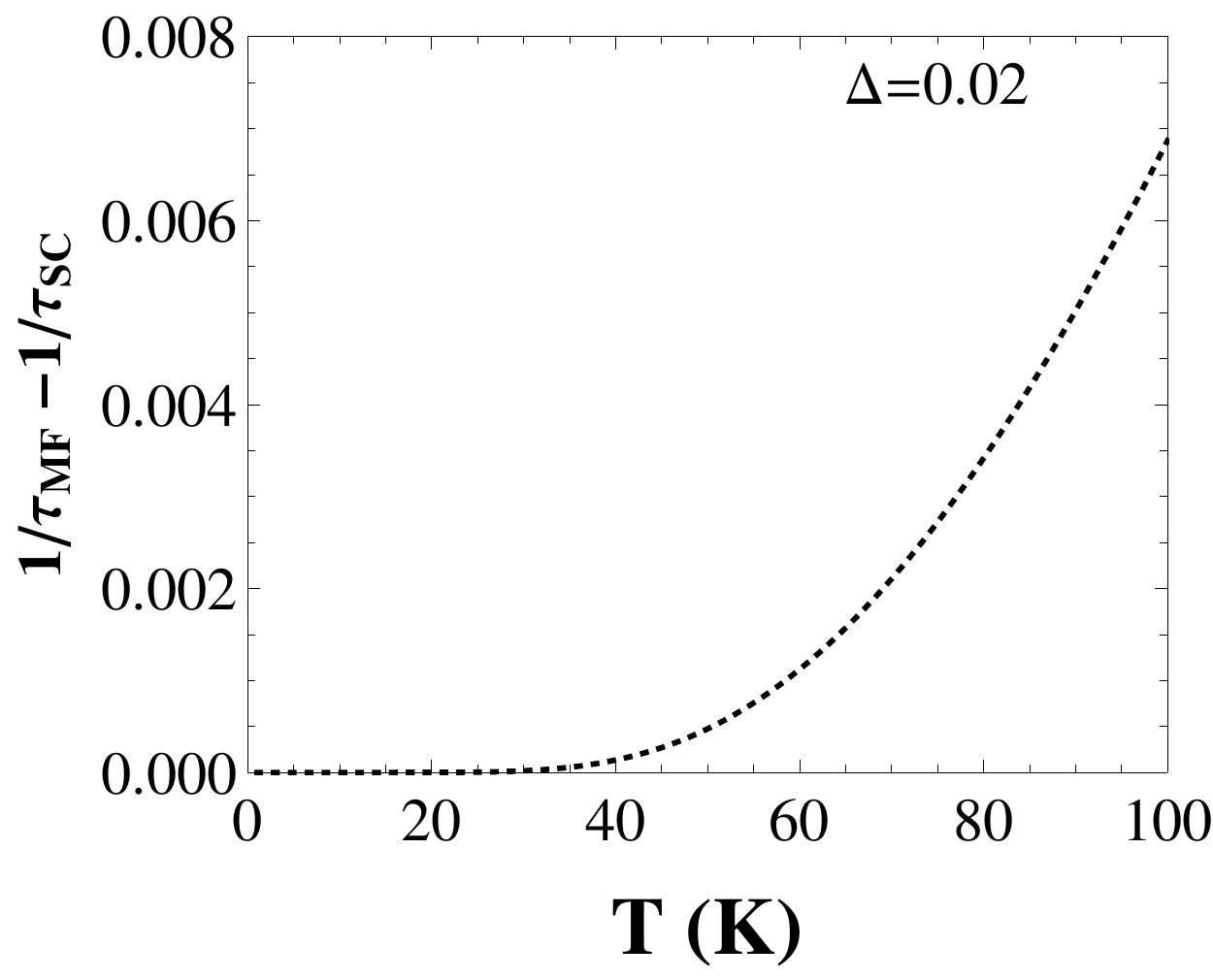}
\caption{ Variation of difference $= \left(1/\tau_{MF}-1/\tau_{SC}\right)$ of dc scattering rate with temperature calculated by two different approaches MF and SC at $\Delta = 0.02$eV.}
\label{tempscales}
\end{figure}

\begin{figure}[htb]
\centering
\hspace{-0cm}
\subfigure[noonleline][]
{\label{self1}\includegraphics[height=30mm,width=40mm]{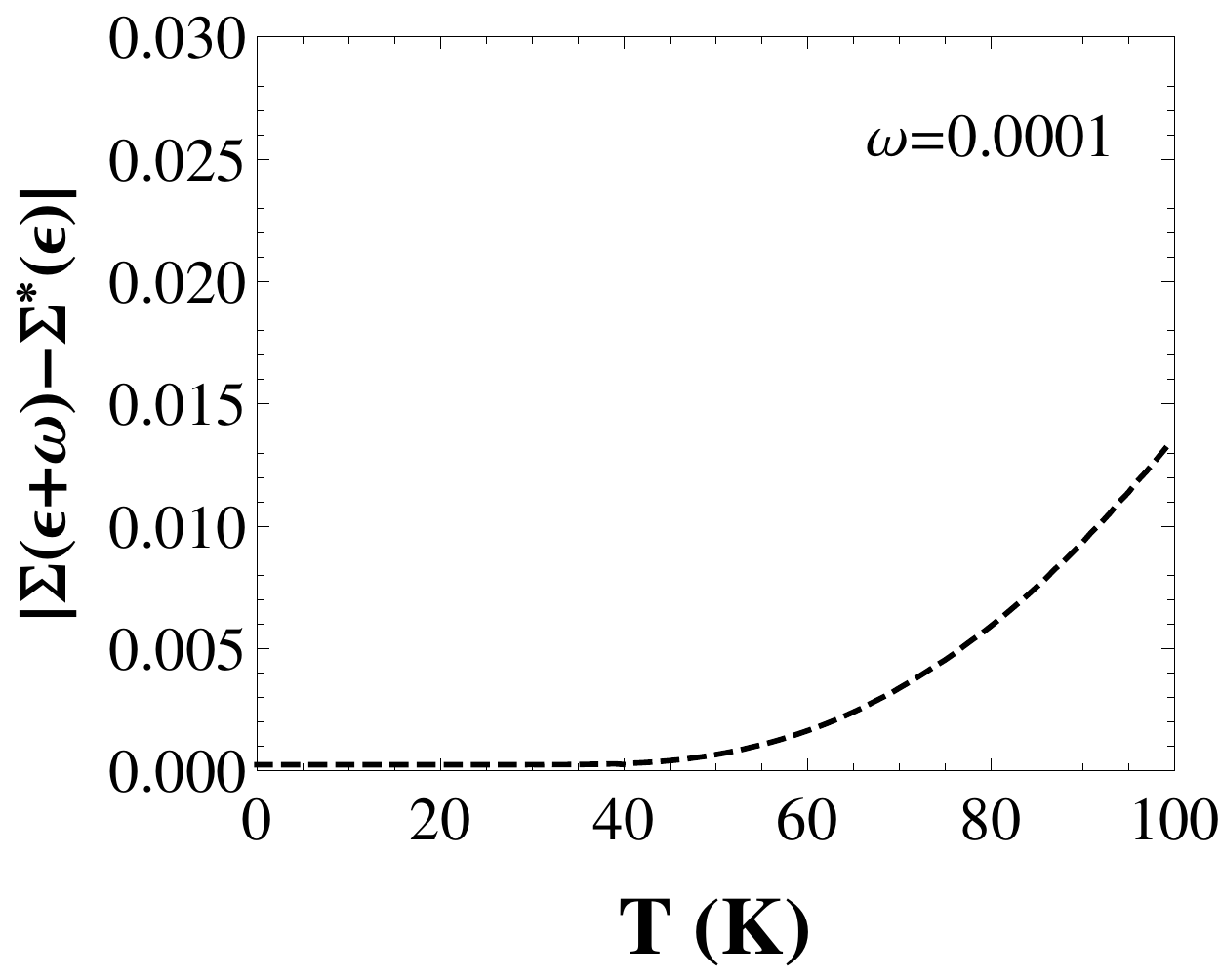}}
\hspace{-0.2cm}
\subfigure[noonleline][]
{\label{self2}\includegraphics[height=30mm,width=40mm]{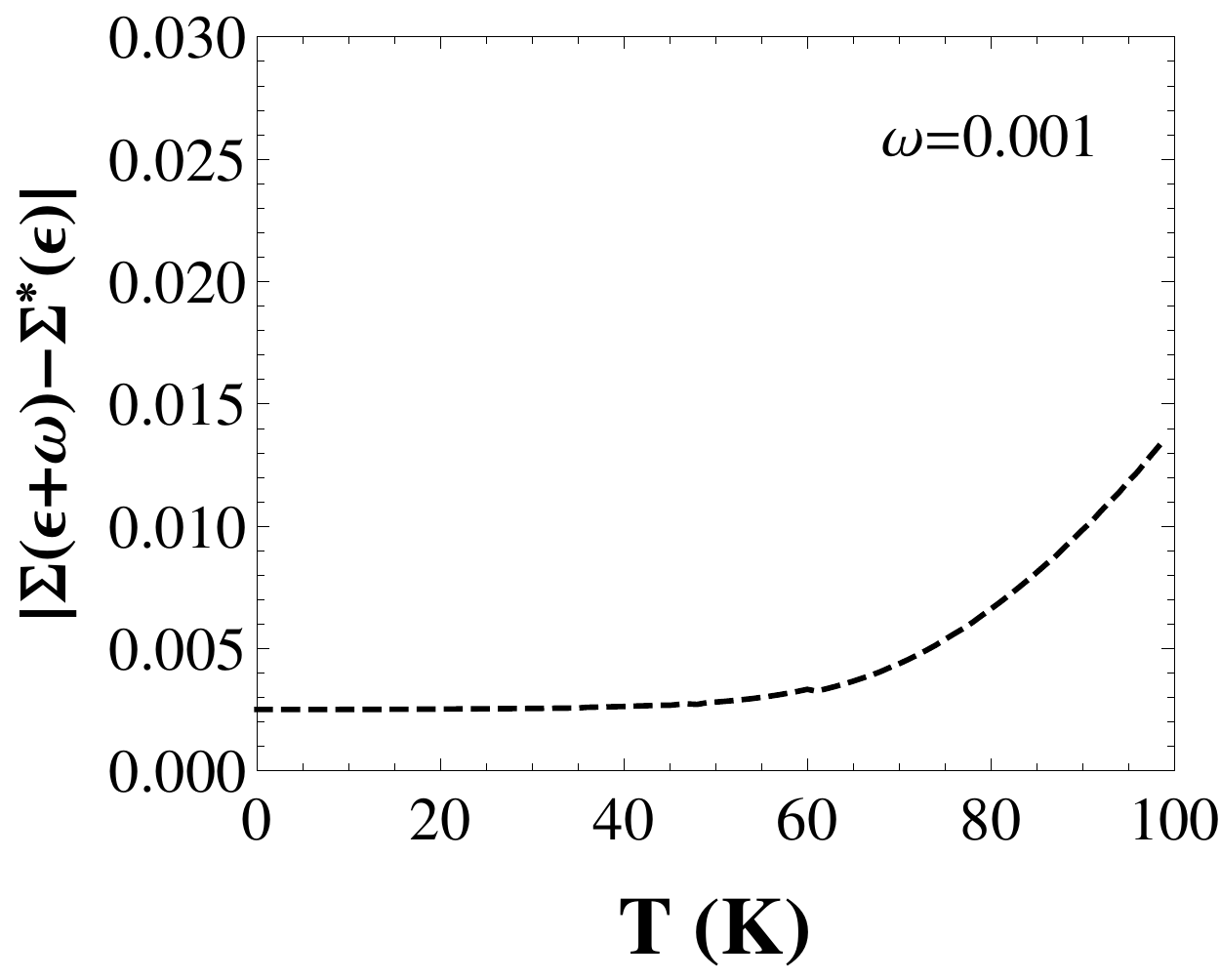}}
\hspace{-0.2cm}
\subfigure[noonleline][]
{\label{self2}\includegraphics[height=30mm,width=40mm]{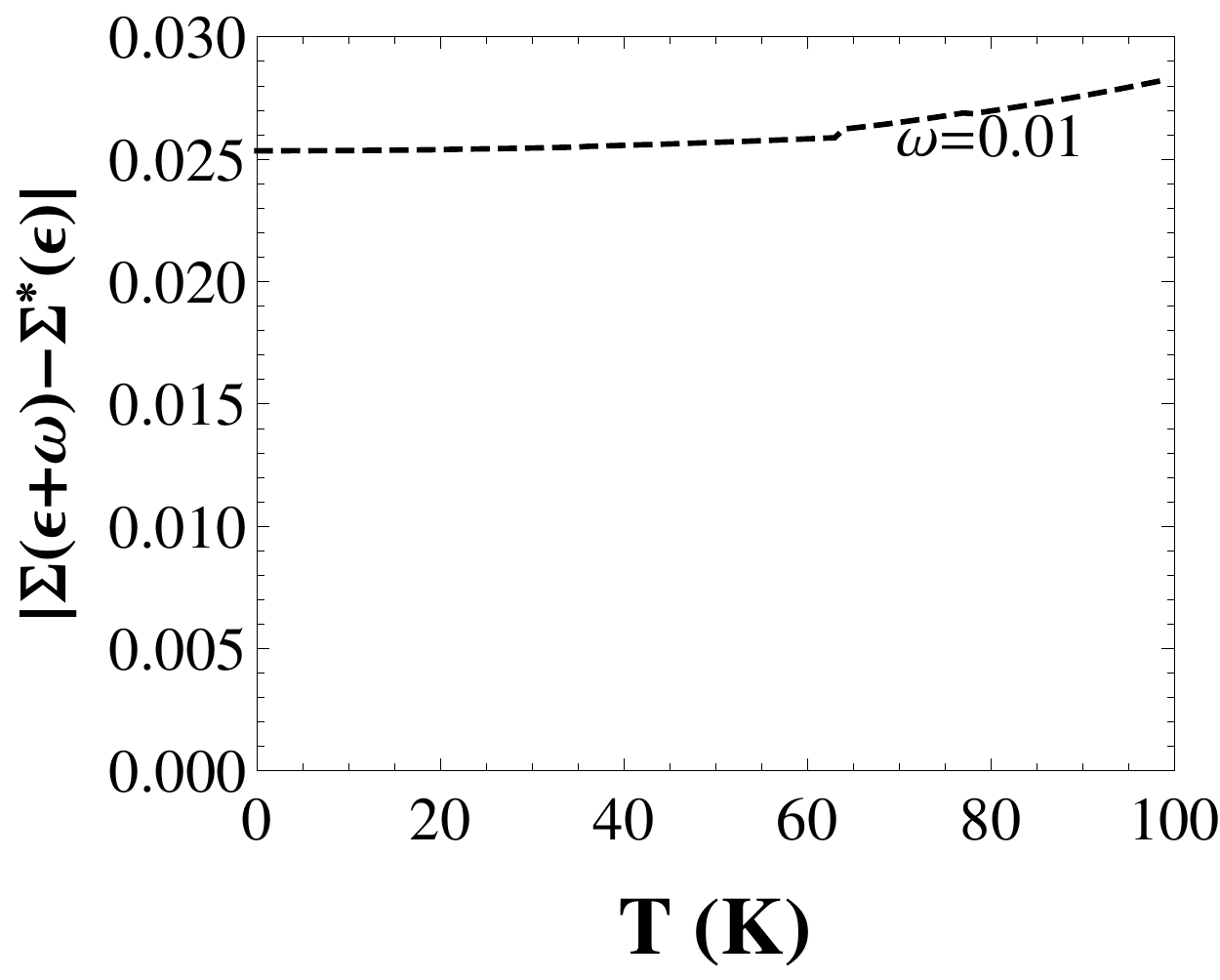}}
\caption{ Plot of $\vert \Sigma(\epsilon+\omega) - \Sigma^{*}(\epsilon)\vert$ with temperature at different frequencies such as (a) $\omega = 0.0001$eV, (b) $\omega = 0.001$eV and (c) $\omega = 0.01$eV . Here $\Sigma(\omega)$ represents the self energy and $*$ corresponds to the conjugate.}
\label{self}
\end{figure}

It is observed that the finite frequency scattering rate using memory function formula is in excellent agrrement with the same obtained from SC formula as shown in fig.(\ref{temp}). This shows that the assumptions made in these two different approaches are consistent at finite frequencies. However, while discussing dc scattering rate, we observe significant discrepancy between the two approaches (fig.(\ref{schm4}) and fig.(\ref{dcconductivity})). To illustrate it further, we have plotted the difference in the magnitudes of scattering rates calculated by both approaches. The difference $\left( 1/\tau_{MF}-\right.$ \\
$\left.1/\tau_{SC} \right)$ at $\Delta=0.02$eV is plotted in fig.(\ref{tempscales}). Here we find that this difference increases with the rise of temperature. The reason behind this difference in the dc case is the assumption made by SC i.e. $\omega >> \vert \Sigma(\epsilon+\omega) - \Sigma^{*}(\epsilon)\vert$ which becomes more severe in high temperature regime. To clarify this fact, we plot the quantity $\vert \Sigma(\epsilon+\omega) - \Sigma^{*}(\epsilon)\vert$ as a function of temperature in fig.(\ref{self}) (where the expression used for $\Sigma(\omega)$ has been given in ref.\cite{sharapov_05}). It shows that the magnitude of the difference of self energy increases with the temperature. This shows the stronger violation of the condition $\omega >> \vert \Sigma(\epsilon+\omega) - \Sigma^{*}(\epsilon)\vert$ in high temperature limit. It implies that SC formalism is not appropriate to study the dc behavior and the disagreement is severe at high temperature, but it is quite reasonable for the finite frequency case.\\
\begin{figure}[htb]
\centering
\hspace{-0cm}
\subfigure[noonleline][]
{\label{validity1}\includegraphics[height=30mm,width=40mm]{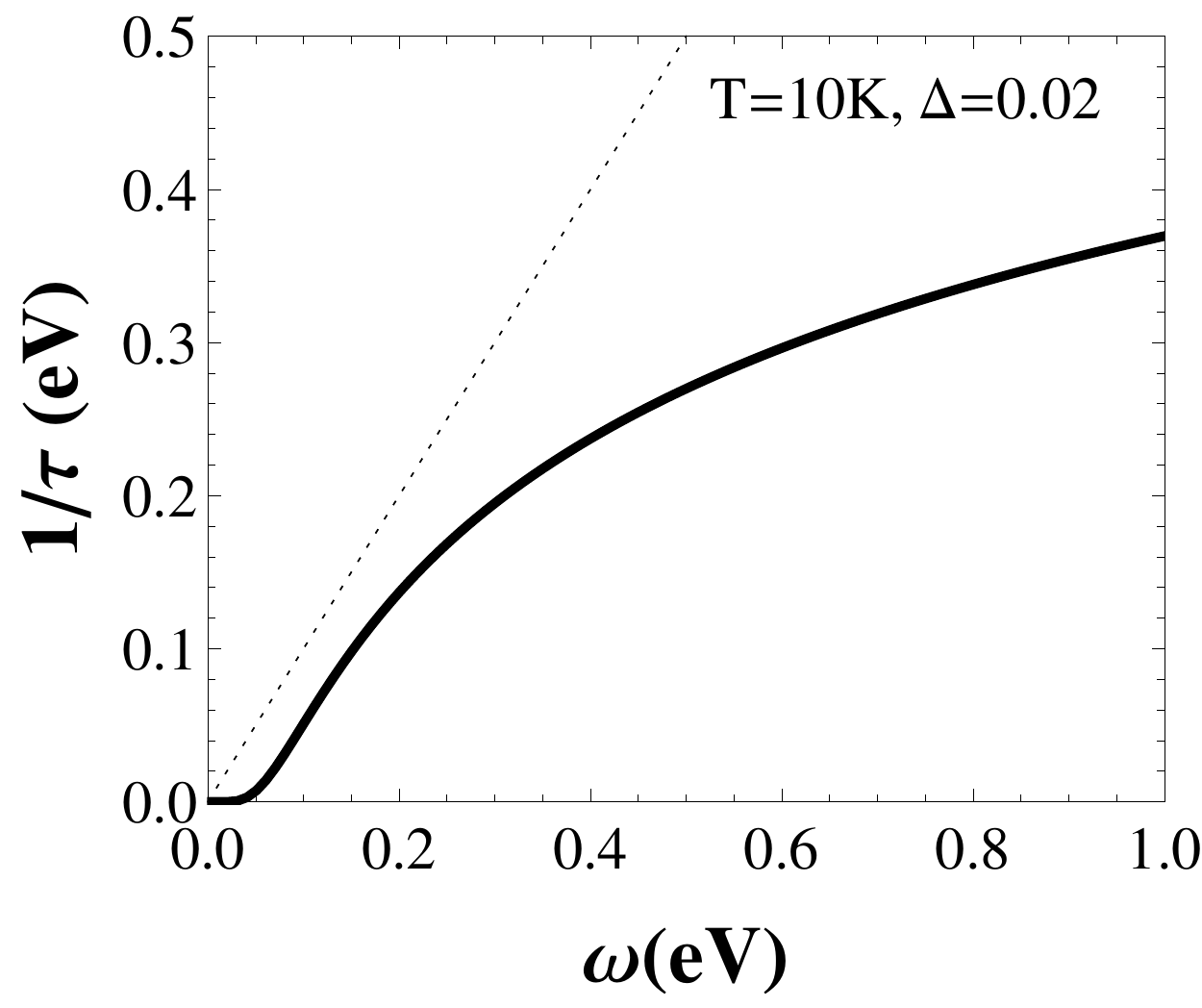}}
\hspace{-0.2cm}
\subfigure[noonleline][]
{\label{validity2}\includegraphics[height=30mm,width=40mm]{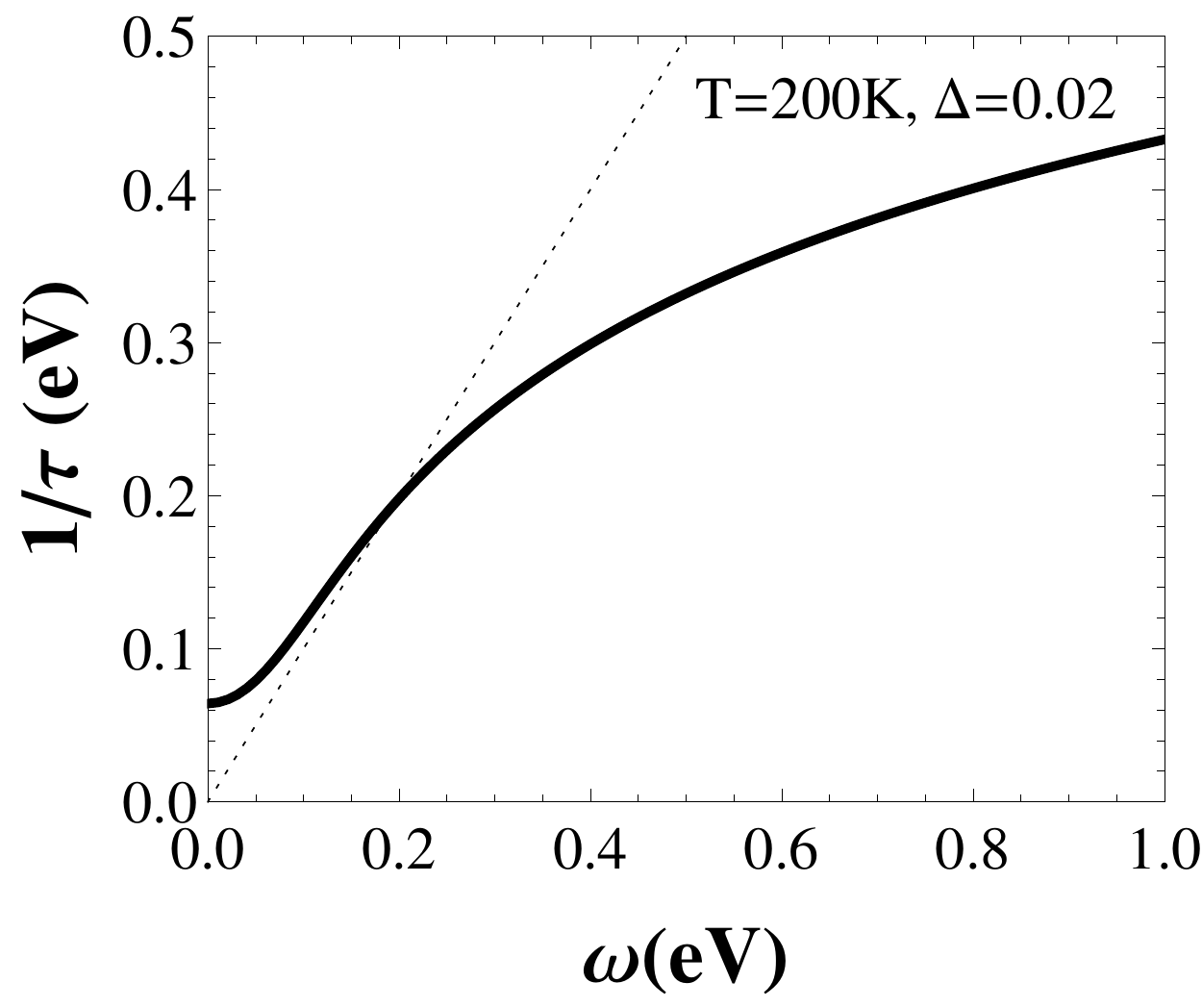}}
\hspace{0cm}
\subfigure[noonleline][]
{\label{validity3}\includegraphics[height=30mm,width=40mm]{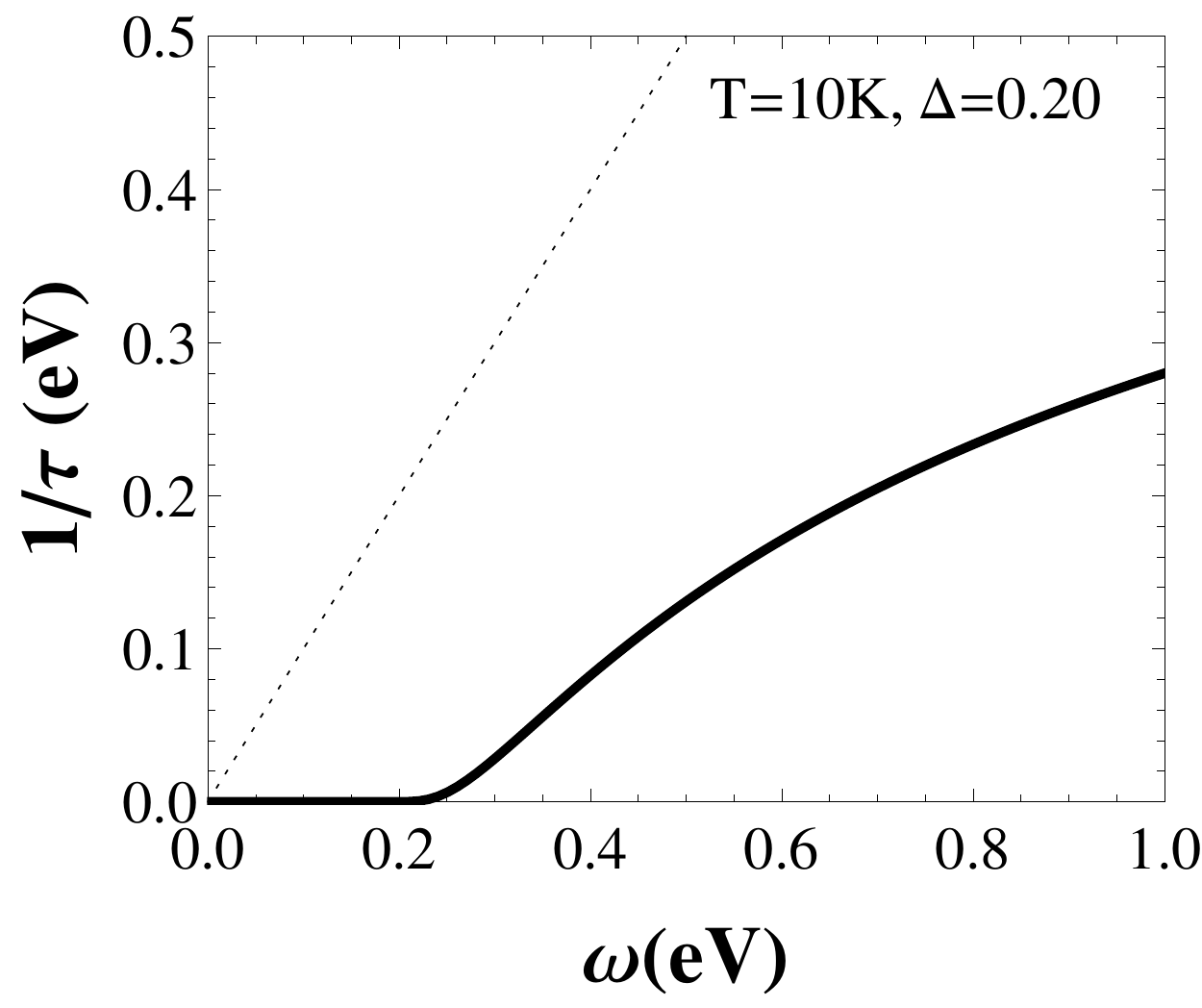}}
\hspace{-0.2cm}
\subfigure[noonleline][]
{\label{validity4}\includegraphics[height=30mm,width=40mm]{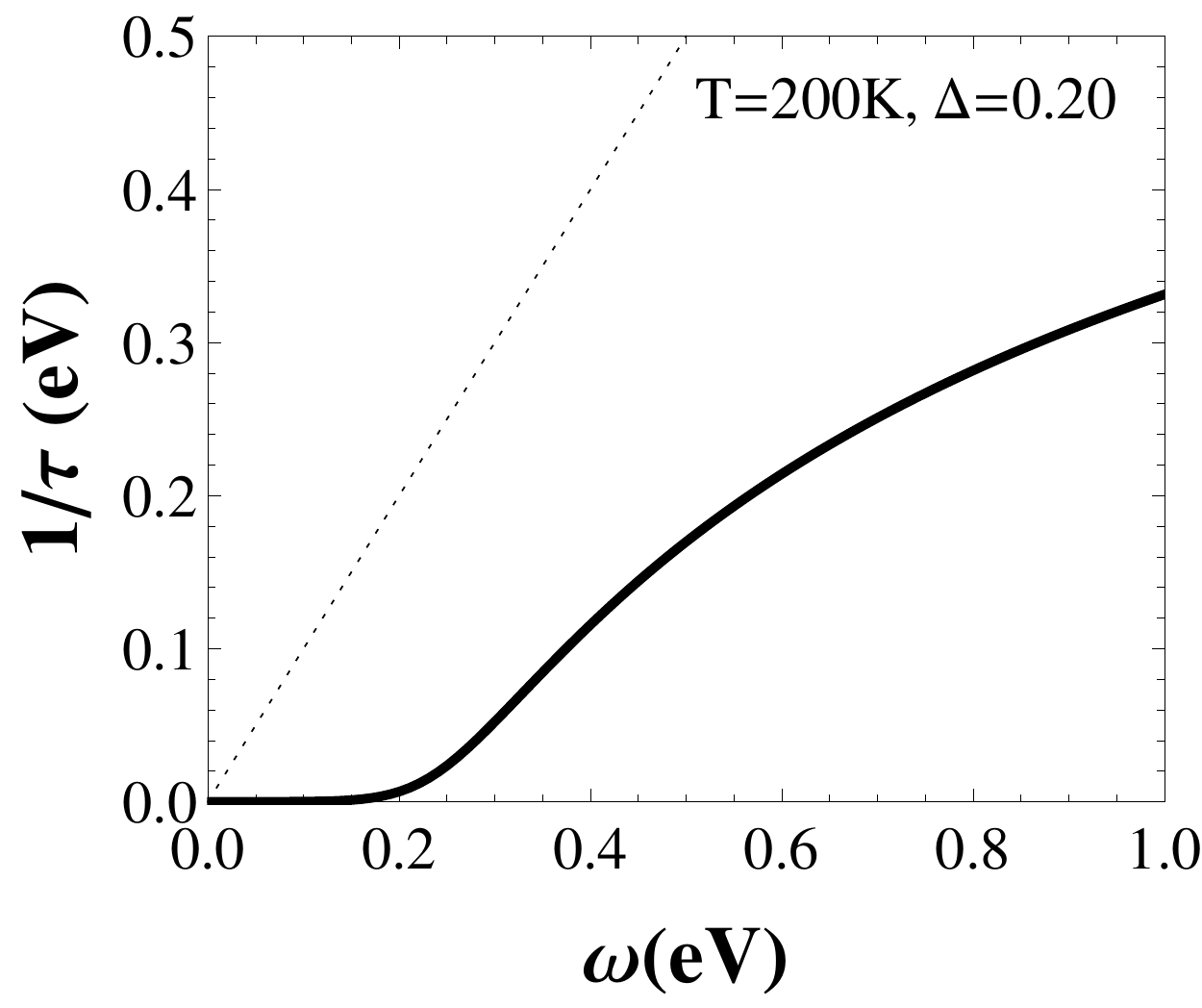}}
\caption{ Solid line: Variation of scattering rate with frequency at different temperature and $\Delta$ (in eV). Dotted line: Plot of $f(\omega)=\omega$ to check the validity of memory function approach.}
\label{validity}
\end{figure}
{\textit{Regime of validity of Memory function:}} It is important to recognize that memory function calculation of $1/\tau(\omega)$ presented here is also perturbative in character and has its own limitations. Here in calculation, we have used the expansion for $M(z)$ which is given below eqn.\ref{conddi} where $M(z)= (z\chi(z)/\chi_{0}) \left( 1
+\chi(z)/\chi_{0}+...\right)$. In the leading order approximation, we took $\chi(z)/\chi_{0} << 1$, which implies that the magnitude of memory function is smaller than $z$ $(M(z)/z = \chi(z)/\chi_{0})$. Thus to ensure this point we have plotted the frequency dependent scattering rate for temperature 10 K and 200 K at $\Delta = 0.02$eV and $0.20$eV in fig.(\ref{validity}) and compared with the linear variation $f(\omega)= \omega $. From fig.(\ref{validity}), we find that our approach is valid for regime where $\Delta >> T$ (as shown in fig.(\ref{validity1}), \ref{validity3} and \ref{validity4}). Thus it is quite suited to study the pseudogap phase of cuprates when $\Delta$ is greater than the $T$. But our approach is not good \textit{in the low frequency regime} to discuss the case where $\Delta \sim T$ (as shown in fig.(\ref{validity2})). Such small gap scenarios occur in spin/charge density systems\cite{gruner_88,gruner_94,gruner_book}.
\section*{Acknowledgement}
\label{sec:5}
The authors are thankful to Nabyendu Das for helpful discussions.

\section*{Authors Contribution}
Both authors contributed equally to the paper.

\appendix
\section{Appendix-I}
\label{Appendix}
The imaginary part of memory function (eqn.\ref{imagmem}) is
\begin{eqnarray} \nonumber
M''(\omega, T)&=&\frac{2\pi}{3} \frac{1}{mN_{e}} \sum_{\textbf{k},\textbf{k}'} \vert D(\textbf{k}-\textbf{k}') \vert ^{2} (\textbf{k}-\textbf{k}')^{2} f'(1-f)n \\ \nonumber
&& \left[ \frac{e^{\omega /T}-1}{\omega} \delta(\epsilon-\epsilon^{\prime}-\omega_{\textbf{k}-\textbf{k}'}+ \omega)\right. \\ 
&&\left.+(\textnormal{terms with} \,  \omega \rightarrow -\omega) \right].
\end{eqnarray}
Converting the summations into energy integrals and inserting the $dq$ integral as before, we have
\begin{eqnarray} \nonumber
M''(\omega, T)&=&\frac{2\pi}{3} \frac{N^{2}}{mN_{e}} \int_{0}^{\infty} dq \int_{-\infty}^{\infty} d\epsilon N(\epsilon) \int_{-\infty}^{\infty} d\epsilon^{\prime} N(\epsilon^{\prime}) \\ \nonumber 
&& \int_{0}^{\pi} d\theta \sin\theta \delta(q-\vert \textbf{k}-\textbf{k}' \vert) \vert D(q) \vert ^{2} q^{2} f'(1-f)n \\ \nonumber
&&  \left[ \frac{e^{\omega /T}-1}{\omega} \delta(\epsilon-\epsilon^{\prime}-\omega_{q}+ \omega)\right. \\
&& \left. +(\textnormal{terms with} \,  \omega \rightarrow -\omega) \right].
\end{eqnarray}
Here the energy dependent density of states $N(\epsilon)$ has been introduced. Thus on solving the integrals over $\epsilon^{\prime}$ and $\theta$, the above equation reduces to
\begin{eqnarray} \nonumber
M''(\omega, T)&=&\frac{2\pi}{3} \frac{N^{2}}{mN_{e}k_{F}^{2}\omega} \int_{0}^{\infty} dq \vert D(q) \vert ^{2} q^{3}  \\ \nonumber
&& \int_{-\infty}^{\infty} d\epsilon N(\epsilon) \frac{e^{\beta(\epsilon-\epsilon_{F})}}{e^{\beta(\epsilon-\epsilon_{F})}+1} \frac{1}{e^{\beta \omega_{q}}-1}  \\ \nonumber
&&   \left[ N(\epsilon-\omega_{q}+\omega) \frac{e^{\beta \omega}-1}{e^{\beta(\epsilon-\epsilon_{F}-\omega_{q}+\omega)}+1}\right. \\ 
&& \left. -N(\epsilon-\omega_{q}-\omega) \frac{e^{-\beta \omega}-1}{e^{\beta(\epsilon-\epsilon_{F}-\omega_{q}-\omega)}+1} \right].
\end{eqnarray}
This is the general expression for the imaginary part of memory function (called as GDS).

\end{document}